\documentclass[twocolumn]{aastex701}
\usepackage{amsmath}

\newcommand{\G}{\Gamma}

\newcommand{\p}{^\prime}
\newcommand{\e}{\epsilon}

\newcommand{\g}{\gamma}

\newcommand{\psim}{\lower.5ex\hbox{$\; \buildrel \propto \over\sim \;$}}
\newcommand{\lbar}{\lower.0ex\hbox{$\; \buildrel{\lower0.0ex \hbox{-}} \over\lambda  \;$}}

\newcommand{\cm}{\mathrm{cm}}

\newcommand{\erg}{\mathrm{erg}}

\newcommand{\MeV}{\mathrm{MeV}}

\newcommand{\s}{\mathrm{s}}

\newcommand{\pc}{\mathrm{pc}}
\newcommand{\kpc}{\mathrm{kpc}}

\newcommand{\yr}{\mathrm{yr}}

\newcommand{\Kelvin}{\mathrm{K}}

\newcommand{\rxj}{RX~J1713.7$-$3946}

\usepackage{graphicx}

\shorttitle{SNR}
\shortauthors{Ziegler et al.}

\begin{document}

\title{Cosmic Ray Electron Evolution in Supernova Remnants:  Log-Parabola Distribution}

\author[0000-0001-8415-4063]{Joshua J. Ziegler}
\affil{National Research Council Research Associate, 
National Academy of Sciences, 
Washington, DC 20001, USA
}
\affil{Resident at U.S.\ Naval Research Laboratory, Washington, DC 20375, USA}

\email{joshua.j.ziegler6.ctr@us.navy.mil}

\author[0000-0001-5941-7933]{Justin D.\ Finke}
\affil{U.S.\ Naval Research Laboratory, 4555 Overlook Ave.\ SW,
        Washington, DC,
        20375-5352, USA \\
         \\
}
\email{justin.d.finke.civ@us.navy.mil}

\author[0000-0002-3362-7040]{J.\ Martin Laming}
\affil{U.S.\ Naval Research Laboratory, 4555 Overlook Ave.\ SW,
        Washington, DC,
        20375-5352, USA \\
         \\
}
\email{j.m.laming.civ@us.navy.mil}

\date{\today}

\begin{abstract}
The shock fronts of supernova remnants (SNRs) are believed to be significant sites of acceleration of cosmic ray particles. Previous researchers have shown that a particle distribution similar to a log-parabola can be generated when particles have an energy-dependent escape.  We explore the acceleration of electrons at SNR shock fronts, and show that modeling this energy-dependent particle escape model can produce spectral energy distributions consistent with observations of two lepton-radiation-dominated SNRs: \rxj\ and SN 1006.  The model includes the evolution of both the electron distribution and photon spectra as a result of the combined effects of the SNR evolution and electron energy loss. The electron-escape energy dependence is quite weak, but the electron distribution and photon spectra turn out to be very sensitive to changes in the electron escape.  
We also explore how sensitive the spectra and electron distributions are to the parameters used in the log-parabola model.
\end{abstract}


\section{Introduction}

A supernova remnant (SNR) is produced when a supernova ejects stellar material {\em en masse} into the surrounding interstellar medium (ISM). As the SNR expands into the ISM, the high speed and density of the ejecta forms a shockwave where the SNR meets the ISM. As particles cross into this shock region, they are accelerated and reflected back in front of (downstream) the shock front, where they can re-enter the shock region. This process can cause some of the particles to be accelerated to very high energies.  When this mechanism is applied to protons and atomic nuclei, it is one of the primary mechanisms by which cosmic rays are believed to emerge \citep[e.g.,][]{jones91,kirk94,hillas05}. 

However, while general features of this Fermi mechanism for acceleration in shock waves are known, there are also several uncertainties around how the process occurs at the SNR-ISM boundary. For example, after each shock acceleration, there is a probability for the accelerated particle to escape the shock region entirely, but it is uncertain how this probability depends on the energy of the particle. A common approach is to assume that the probability of escape is independent of particle energy, which leads to an integrated distribution of accelerated particles whose energy follows a power-law \citep[e.g.,][]{bell78_paper1}.

Once particles have been accelerated, they emit that energy as radiation. Electrons emit photons through the synchrotron, bremsstrahlung, and inverse-Compton processes, which produce radiation throughout the electromagnetic spectrum; and protons and ions can produce $\g$-rays through pion production and subsequent decay \citep[e.g.,][]{sturner97}.  This broadband electromagnetic radiation can then be probed by different observatories, operating in a wide range of frequencies, and compared with predictions based upon the power-law injection particle distribution.  Non-linear effects can modify the accelerated particle spectrum so that it deviates from a power-law \citep[e.g.,][]{ellison84}.

In this paper, we focus on electron acceleration and radiation; inclusion of accelerated protons will be left for future work.  Some SNRs, such as \rxj\ and SN 1006, are thought to emit essentially all their observed radiation through leptonic (electron) processes, although a subdominant hadronic component may be possible.  Previously, \citet{finke12} attempted to model multiwavelength observations of \rxj\ with a leptonic model that included accelerated power-law electron distributions, and subsequent cooling.  They were not able to reproduce the multiwavelength spectral energy distribution (SED) of this remnant with a single zone model; they required emission from a second zone of compact knots. Indeed, fully explaining the broadband SED, including the LAT data, is quite challenging.  \citet{zeng16} and \citet{cristofari21} modeled \rxj\ were able to reproduce much of the soft LAT $\g$-rays with a hadronic component, while the higher energy $\g$-rays were explained by inverse Compton scattering.  \citet{abdalla18} used a broken power-law electron distribution to fully explain the $\g$-ray SED.   \citet{gabici14} modeled the $\g$-ray emission from \rxj\ as decay of pions from hadronic interactions with gas clumps in the SNR.

We also explore modeling of SN 1006, an SNR that is also usually modeled with a fully leptonic origin of its radiation from, although some authors do model it with a hadronic component \citep{acero15,xing16,winner20,lemoine-goumard25}. Accelerated particle distributions are typically modeled as power laws with exponential cutoffs at high energies. The hadronic contribution to the gamma ray spectrum of the northeast lobe of the SN 1006 remnant is significantly subdominant, so the northeast lobe can be modeled almost purely leptonically. The southwestern lobe meanwhile is colliding with a dense cloud of interstellar gas, increasing the number of potential proton-proton collision centers and enhancing the hadronic contribution to the gamma ray spectrum.

For an entirely different sort of astrophysical object, blazars,\footnote{Active Galactic nuclei with relativistic jets oriented to our line of sight.} X-ray spectra were found to be fit by log-parabola models.  This led \citet{massaro04} to propose an electron acceleration model that had an energy-dependent particle escape term that could lead to log-parabola electron distributions, which would then produce log-parabola X-ray spectra through synchrotron emission.  This model for accelerated electrons was applied by \citet{fraschetti17} to the acceleration of particles in the Crab Pulsar Wind Nebula, and shown to provide a good reproduction of the source's SED.  Here we apply the model to \rxj\ and SN 1006.  Our motivation is to provide a better model with a single zone for \rxj\ compared to the one by \citet{finke12}; and to gain a better understanding of the particle acceleration mechanism in SNRs.

This paper is laid out as follows. In Section~\ref{Formalism}, we discuss in more detail the models we use for the SNR, acceleration mechanism, evolution of electrons after being accelerated, and the radiation spectra emitted. We then discuss how the evolving SNR environment produces changes in the electron distribution and observable spectra in Section~\ref{Evolution}. We discuss the models  that we find to suitably approximate observed spectra of SNR \rxj\ and SN 1006 in Sections~\ref{RXJ1713} and \ref{SN1006}, respectively. Finally in Section~\ref{Discussion}, we conclude with a discussion of our results.

\section{Formalism}\label{Formalism}

We primarily follow the model by \citet{sturner97} (see also \citet{berrington03}), with some modifications for a log-parabola electron distribution.  We neglect proton acceleration and radiation in this model, which we leave for future work.  

\subsection{SNR Dynamics}
As a SNR expands, it evolves through three distinct phases.
In the free expansion phase, the radius $R$ and expansion speed $v$ of the remnant as a function of time $t$ are given by
\begin{flalign}
R = v_0 t\ ,
\end{flalign}
and
\begin{flalign}
v = v_0\ ,
\end{flalign}
respectively, where $v_0$ denotes the initial expansion speed of the remnant, which we will assume is nonrelativistic. This ends at the Sedov time, when the mass of the interstellar medium swept up by the SNR is comparable to the initial ejected mass of the supernova. For a supernova that ejects $M_{ej}$ into an interstellar medium with density $n_{\rm{ISM}}$ and average atomic mass $\mu\approx 1.4$, the Sedov time is given by
\begin{flalign}
t_{\rm Sed} = 
\frac{1}{v_0}\left[\frac{3M_{\rm ej}}{4\pi \mu m_p n_{\rm ISM}}\right]^{1/3}\ .
\end{flalign}
After the Sedov time, the SNR enters the Sedov phase, where its radius and expansion speed are given by
\begin{flalign}
R = 2.5(v_0 t_{\rm Sed}) \left[\left(\frac{t}{t_{\rm Sed}}\right)^{0.4} - 0.6\right]\ ,
\end{flalign}
and
\begin{flalign}
v = v_0 \left(\frac{t}{t_{\rm Sed}}\right)^{-0.6} \ ,
\end{flalign}
respectively.
This phase lasts until time
\begin{flalign}
t_{\rm rad} = 2.9\times10^4 \left(\frac{E_0}{10^{51}\ \erg}\right)^{0.24}
n_{\rm ISM}^{-0.52}\ ,
\end{flalign}
where the initial supernova explosion energy $E_0=\frac{1}{2} M_{ej} v_0^2$. After $t_{\rm rad}$, the SNR radius
\begin{flalign}
R & = 2.5 (v_0t_{\rm Sed}) \Biggr\{1.29\left(\frac{t_{\rm rad}}{t_{\rm Sed}}\right)^{0.4}
\nonumber \\ & \times
\left[\left(\frac{t}{t_{\rm rad}}\right)^{0.31}-0.225\right]  -0.6 \Biggr\}\ ,
\end{flalign}
and velocity
\begin{flalign}
v = v_0\left(\frac{t_{\rm Sed}}{t_{\rm rad}}\right)^{0.6}
\left(\frac{t}{t_{\rm rad}}\right)^{-0.69}\ .
\end{flalign}
The phase where $t>t_{\rm rad}$ is not relevant to the SNRs explored here, but is included for completeness.

The shock will have a compression ratio
\begin{flalign}
\label{chi}
\chi = \frac{\g_{\rm ad}+1}{\g_{\rm ad}-1 + 2/\mathcal{M}^2}\ ,
\end{flalign}
where the adiabatic index $\g_{\rm ad}=5/3$ for a nonrelativistic monatomic ideal gas, the Mach number of the shock
\begin{flalign}
\mathcal{M} = \frac{v}{c_s}
\end{flalign}
where the sound speed 
\begin{flalign}
c_s = \sqrt{\frac{k_bT}{m_p\mu}}\ ,
\end{flalign}
$k_B\approx1.38\times10^{-16}\ \erg\ \Kelvin^{-1}$ is the Boltzmann Constant, and we take the temperature of the ISM $T=10^4\ \Kelvin$.

\subsection{Particle Acceleration}
\label{sec:particleaccel}

We primarily follow \citet{massaro04} to derive the distribution of accelerated electrons, with a small modification. \citet{massaro04} perform this analysis using an acceleration that directly depends on the Lorentz factor $\g$. However, a similar analysis can be performed where the analysis is directly dependent on the momentum $p$. Deviations between these two approaches are insignificant at high energy, while observations at low energy seem to favor the momentum-based approach \citep[e.g.,][]{dermer12}. 

Electrons begin with some initial momentum $p_0$.  Each time an electron travels through the SNR shock front, it increases its momentum by a constant factor $\varepsilon$,
so that the momentum of an electron after crossing through the shock for the $i$th time is
\begin{flalign}
p_i = \varepsilon p_{i-1}\ ,
\end{flalign}
or, after $n$ times crossing through the shock
\begin{flalign}
\label{gn}
p_n = \varepsilon^n p_0\ .
\end{flalign}
We assume the probability that an electron is accelerated and avoids escape is
\begin{equation}
P_i = \begin{cases}
P_0 & p_i < p_c \\
g\left(p_ic\right)^{-q} & p_c < p_i
\end{cases}
\end{equation}
where $g$, $q$, and $p_c$ are constants, and $P_0 = g\left(p_cc\right)^{-q}\le 1$.  Unlike the version by \citet{massaro04}, this has constant probability at low energy, which allows the model to avoid probabilities greater than unity.  We take the step associated with $p_c$ to be $i=i_c$, so that $p_c = \varepsilon^{i_c}p_0$.  The number of electrons after step $i$ is
\begin{flalign}
N_i = P_i N_{i-1}\ .
\end{flalign}
The electron distribution is found as follows.  For $p_i<p_c$, one has
\begin{flalign}
N_n = N_0 \prod_{i=0}^{n-1}P_i = P_0^n N_0
\end{flalign}
where $N_0$ is the number of initial electrons at $p_0$.  Following Equation (\ref{gn}), one has
\begin{flalign}
n = \frac{\ln(p/p_0)}{\ln\varepsilon}\ ,
\end{flalign}
so that
\begin{flalign}
Q(>pc) \equiv N_n = N_0\left(\frac{pc}{p_0c}\right)^{-(s\p-1)}\ ,
\end{flalign}
where
\begin{flalign}
s\p = 1 - \frac{\ln P_0}{\ln\varepsilon}\ .
\end{flalign}
For $p_c<p_i$, 
\begin{flalign}
N_n & = N_0 \prod_{i=0}^{n-1}P_i = N_0 P_0^{i_c} \prod_{i=i_c}^{n-1}g\left(p_ic\right)^{-q}
\nonumber \\
 & = N_0P_0^{i_c}\frac{g^{n-i_c}}{\prod_{i=0}^{n-i_c-1}\left(p_cc \varepsilon^{i}\right)^q}
 \nonumber \\
 & = N_0P_0^{i_c}\left(p_cc\right)^{-q(n-i_c)}\varepsilon^{-q(n-i_c)(n-i_c-1)/2}\ .
\end{flalign}
Since 
\begin{flalign}
n-i_c = \frac{\ln(p/p_c)}{\ln\varepsilon}\ ,
\end{flalign}
one gets
\begin{flalign}
Q(>pc) & \equiv N_n 
\nonumber \\
& = N_0 \left(\frac{p_cc}{p_0c}\right)^{-(s\p-1)}
\left(\frac{pc}{p_cc}\right)^{-[s-1+r\ln(p/p_c)]}\ ,
\end{flalign}
where
\begin{flalign}
s & = \frac{2-q}{2} - \frac{\ln P_0}{\ln\varepsilon}  \ ,
\\
r & = \frac{q}{2\ln\varepsilon}\ .
\end{flalign}
From standard test particle shock acceleration theory \citep[e.g.,][]{bell78_paper1}, 
\begin{flalign}
\varepsilon = 1 + \frac{4}{3}\frac{v}{c}\left(1-1/\chi\right)  \ .
\end{flalign}

The accelerated particle distribution is typically
written as
\begin{flalign}
Q(pc) = \frac{d}{d(pc)} Q(>pc)\ .
\end{flalign}
We re-write this with normalization constant, $Q_0(t)$, that evolves with time $t$ as the remnant expands, giving
\begin{eqnarray}
Q(pc, t)  = Q_0(t) \frac{|s\p-1|}{p_0c}
\left(\frac{p}{p_0}\right)^{-s\p} 
\end{eqnarray}
for $p\le p_c$, and
\begin{eqnarray}
Q(pc, t) & =  Q_0(t) 
\frac{|s-1+2r\ln(p/p_c)|}{p_0c}
\nonumber \\ & \times
\left(\frac{p_c}{p_0}\right)^{-s\p}
\left(\frac{p}{p_c}\right)^{-[s+r\ln(p/p_c)]}
\end{eqnarray}
for $p>p_c$.  Note that, while $Q(>pc,t)$ is continuous, $Q(pc,t)$ has a discontinuity at $p=p_c$ that will be small for small values of $q$. 
In the Appendix we discuss the choice of functional form for the probability and acceleration model and its accuracy.

\subsubsection{Particle Acceleration in Terms of Kinetic Energy}

We wish to write the electron particle source function $Q(pc)$, in terms of kinetic energy, 
$$
Q(K, t) = \frac{dQ(>K)}{dt dV dK} \ ,
$$
where the the momentum is $p=\g\beta mc$, $p_0=\g_0 \beta_0 mc$, and the kinetic energy is $K=(\g-1)mc^2$.  It is useful to note
\begin{flalign}
\g = \frac{K}{mc^2}+1 = \frac{1}{mc^2}\left( K + mc^2\right)\ ,
\end{flalign}
\begin{flalign}
\beta = \frac{\sqrt{K(K+2mc^2)}}{K+mc^2}\ ,
\end{flalign}
and
\begin{flalign}
pc = \sqrt{K(K+2mc^2)}\ .
\end{flalign}
Also, 
\begin{flalign}
\int d(pc) Q(pc) = \int dK Q(K)\ ,
\end{flalign}
which implies
$$
Q(K, t) = Q(pc, t)\frac{dK}{d(pc)} = \frac{Q(pc, t)}{\beta}\ .
$$
For $p\le p_c$,
\begin{flalign}
Q(K, t) & = \frac{Q_0(t)|s\p-1| (K+mc^2)}{\sqrt{K(K+2mc^2)K_0(K_0+2mc^2)}}
\nonumber \\ & \times
\left( \frac{K(K+2mc^2)}{K_0(K_0+2mc^2)}\right)^{-s\p/2}\ .
\end{flalign}
For $p>p_c$, 
\begin{flalign}
Q(K, t) & = \frac{Q_0(t) (K+mc^2)}{\sqrt{K(K+2mc^2)K_0(K_0+2mc^2)}}\nonumber\\
&\times\left( \frac{K_c(K_c+2mc^2)}{K_0(K_0+2mc^2)}\right)^{-s\p/2}
\nonumber \\ & \times
\left( \frac{K(K+2mc^2)}{K_c(K_c+2mc^2)}\right)
^{-\frac{s}{2} - \frac{r}{2}\ln\left(\frac{K(K+2mc^2)}{K_c(K_c+2mc^2)}\right)}
\nonumber \\
& \times \left|s - 1 + r\ln\left(\frac{K(K+2mc^2)}{K_c(K_c+2mc^2)}\right)\right| \nonumber\\ &\times 
e^{-K/K_{max}}\ .
\label{eqn:Q(K)}
\end{flalign}
In Equation~(\ref{eqn:Q(K)}), we have included an exponential cutoff at high energies in the electron source function. Assuming the Larmor radius of the accelerated electrons remains smaller than the SNR thickness (approximately 10\% the SNR radius), $K_{\max}$ is the lowest of two possibilities:
\begin{flalign}
K_{\max,1} = 100\ \int_0^t dt\p\frac{B v_8^2(t\p)}{fR_J}~\MeV\
\label{eqn:kmax1}
\end{flalign}
and
\begin{flalign}
K_{\max,2} = 2\times10^{5}\ (fR_JB)^{-0.5}v_8~\MeV\ .
\label{eqn:kmax2}
\end{flalign}

 Here $f\approx10$ is the Larmor diffusion factor, $R_J\approx 1$ relates to the orientation between the magnetic field (with magnitude $B$) and shock front, and $v_8$ is the expansion speed of the SNR in units $10^8\ \cm\ \s^{-1}$. 

The first, $K_{\max,1}$ comes from integrating the acceleration rate over the lifetime of the remnant \citep{reynolds95,reynolds96}.  The particles will diffuse across the shock front multiple times, and this ensures they have enough time to reach the highest energies.  The second, $K_{\max,2}$ comes about from the constraint that the particle acceleration timescale must be less than the synchrotron energy loss timescale.  Generally, $K_{\max,1}<K_{\max,2}$ at early times, as the electrons have not had time to be accelerated to the highest energies, so that the diffusion time will set the maximum energy of the electrons;
at late times, $K_{\max,2}<K_{\max,1}$, so that the radiative losses will constrain the maximum energy.

The source function $Q(K, t)$ is normalized at each time $t$ so that 
\begin{flalign}
    \int_{K_{min}}^{K_{max}} dK KQ(K, t) = \frac{\eta_e}{2}A (t)\mu m_p v(t)^3 n_{ISM}\ ,
\end{flalign}
where $A(t)=4\pi R(t)^2$ is the area of the shock front, $v(t)$ is the SNR expansion speed, and $n_{ISM}$ is the number density of the ISM. The parameter $\eta_e$ is an efficiency parameter that measures how much of the energy of the SNR goes into accelerating electrons.   Performing the integral over energy gives the normalization constant $Q_0(t)$.

\subsection{Electron Evolution}
The electron distribution, $N(K;t)$ is given by the solution to a Fokker-Planck Equation, 
\begin{flalign}
\label{eqn:fokker-planck}
\frac{\partial N(K;t)}{\partial t} & = 
\frac{1}{2}\frac{\partial^2}{\partial K^2}[D(K,t) N(K;t)] 
\nonumber \\ & 
- \frac{\partial}{\partial K} \{ \dot K_{\rm tot}(K,t) N(K;t)\}
\nonumber \\ & 
+ Q(K,t)\ .
\end{flalign}
This is solved for $N(K;t)$ numerically with a Crank-Nicholson finite differencing scheme \citep[e.g.,][]{press92}.  In this equation, $Q(K,t)$ is included as described in Section \ref{sec:particleaccel}.

Following \citet{sturner97,berrington03}, we calculate $\dot{K}_{\rm tot}(K,t)$ by considering five mechanisms by which electrons lose energy: Coulomb interactions between accelerated electrons and the thermal protons in the ISM; bremsstrahlung radiation from accelerated electrons interacting with protons in the ISM; synchrotron radiation; inverse Compton scattering of low-energy background photons off of high energy electrons; and adiabatic energy loss due to the expansion of the SNR.  We also include a diffusion coefficient, $D(K,t)$, from Coulomb interactions.  The synchrotron emission includes the magnetic field, modified by the compression factor.  The Compton scattering is of radiation fields from the cosmic background radiation (CMB), and the background optical and infrared radiation from the Galaxy.  All of these are described in detail in \citet{sturner97}.

\subsection{Radiation}

Once $N(K;t)$ is computed from Equation (\ref{eqn:fokker-planck}), this nonthermal electron distribution can be used to compute the radiative emission as a function of time $t$.  We include radiation by synchrotron, Bremsstrahlung, and inverse Compton scattering; see \citet{sturner97} for details.  Bremsstrahlung is generated by the nonthermal electrons interacting with the thermal ISM with density $n_{\rm ISM}$.  We calculate the nonthermal leptonic SEDs of an SNR by summing the flux contributions from these three sources.

\section{SNR Evolution}
\label{Evolution}

\begin{figure}
    \includegraphics[width=\columnwidth]{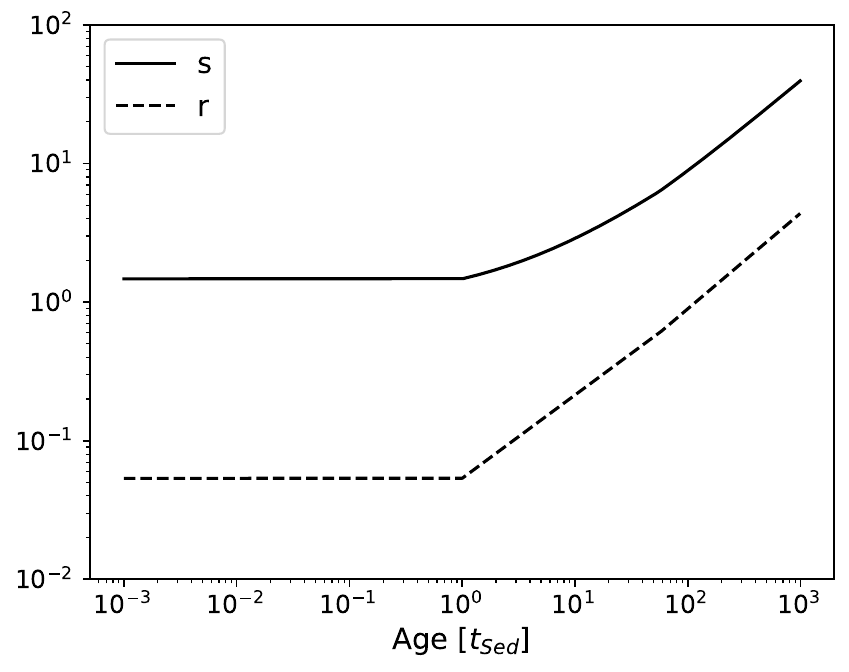}
    \caption{Evolution of the macroscopic acceleration parameters $r$ and $s$ for the model described in Table~\ref{tab:rxj1713fitparams}. Prior to the Sedov time ($t_{\rm Sed}=1)$, both parameters are constant, but as the SNR slows down, electrons are accelerated by a smaller fraction during each shock-crossing, and the parameters $s$ and $r$ increase. This also leads to the slight change in the evolution at time $t_{\rm Rad}\approx60 t_{\rm Sed}$.
    }
    \label{fig:rxj1713rsevol}
\end{figure}

We assume that the parameters $q$ and $g$, as well as the cutoff energy $K_c$ and characteristic energy $K_0$ are fixed throughout the lifetime of an SNR, but may vary between different SNRs. 
Making this assumption implies that the values of the macroscopic quantities $r$ and $s$ evolve with time. We plot this evolution in Figure~\ref{fig:rxj1713rsevol}. To produce this plot, we have used the parameters derived below and listed in Table~\ref{tab:rxj1713fitparams} 
in order to ensure something physically realistic, but the conclusions can be extended to other parameter values. Before the Sedov time, the speed and compression ratio of the remnant are constant, and so the values of $r$ and $s$ are constant. For $t>t_{\rm Sed}$, in the Sedov phase, the remnant begins to slow down, the Mach number decreases, and the compression ratio (Equation (\ref{chi})) decreases, leading to smaller $\varepsilon$ and larger values of $r$ and $s$.  While this increase is slow on human-observable timescales, it suggests that the electron source functions $Q(K,t)$ of old SNRs may generally be expected to be steeper than those of young SNRs.

\begin{table}
\caption{Model parameters for \rxj.} 
\centering
\begin{tabular}{l c c}
\hline
\hline
Parameter & Symbol & Model\\
\hline
\multicolumn{3}{c}{Parameters fixed by observation}\\
\hline
Age [yr] & $t$ & 1620\\
Distance [kpc] & $d$ & $1.0$\\
\hline
\multicolumn{3}{c}{Free parameters constrained by observation}\\
\hline
Initial Mass [M$_\odot$] & M$_{ej}$ & 5.5\\
Initial Energy [$10^{51}$ erg] & $E_0$ & 1.1\\
Magnetic field [$\mu$G] & $B$ & 11.4\\
ISM Density [cm$^{-3}$] & $n_{ISM}$ & 0.175 \\
\hline
\multicolumn{3}{c}{Free parameters}\\
\hline
Acceleration scaling parameter & $q$ & 0.0016 \\
Acceleration normalization [MeV/c] & $g$ & 1.004 \\
Log-Parabola electron energy [MeV] & $K_0$ & $10$\\
Log-Parabola cutoff energy [MeV] & $K_c$ & $10^3$\\
Electron acceleration efficiency & $\eta_e$ & $3.7\times 10^{-4}$\\
\hline
\multicolumn{3}{c}{Derived parameters}\\
\hline
Low-energy probability of acceleration & $P_0$ & 0.993 \\
Initial velocity [cm s$^{-1}$] & $v_0$ & $4.5\times10^{8}$ \\
Sedov time [yr] & $t_{\rm Sed}$ & 1310\\
Present radius [pc]  & $R_t$ & 7.4\\
Present velocity [cm s$^{-1}$] & $v_t$ & $4.0\times10^8$\\
\hline
\end{tabular}
\label{tab:rxj1713fitparams}
\end{table}

\begin{figure*}
    \includegraphics[width=\columnwidth]{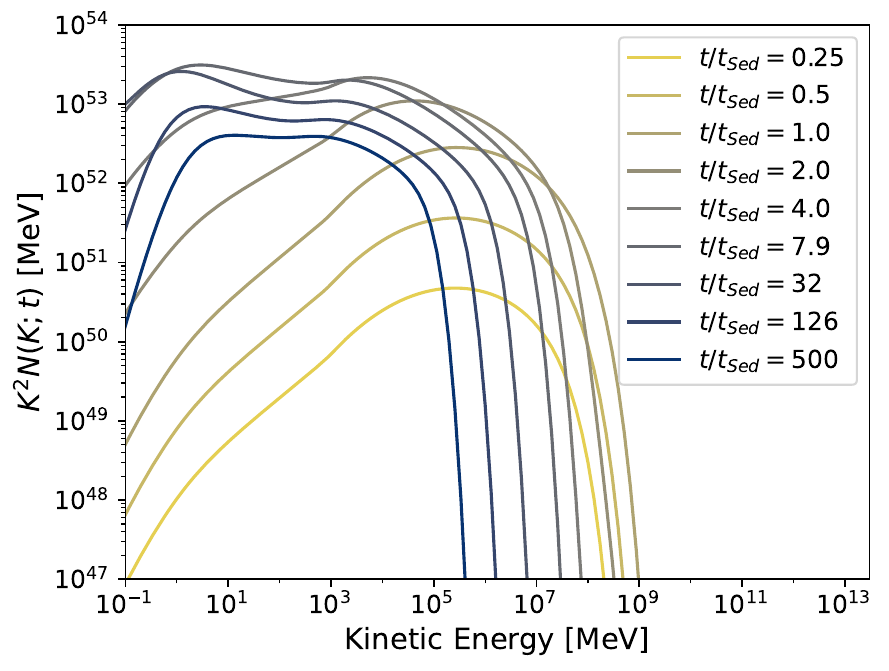}
    \includegraphics[width=\columnwidth]{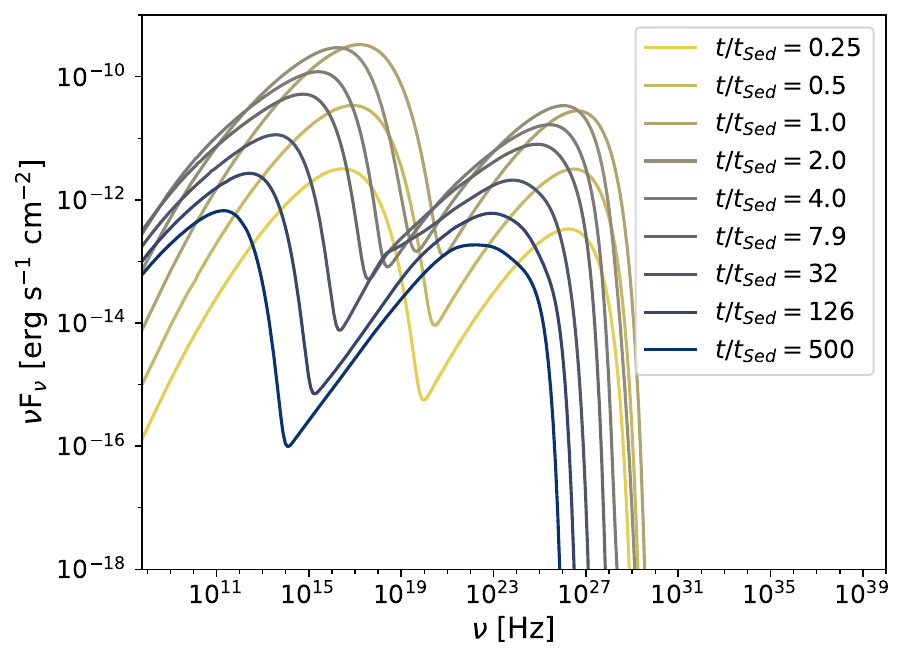}
    \caption{Evolution of electron distribution (left) and photon spectrum (right) over time for the model described in Table~\protect{\ref{tab:rxj1713fitparams}}.
    } 
    \label{fig:rxj1713evol}
\end{figure*}

The evolution of the quantities $s$ and $r$ as well as the evolution of the normalization $Q_0(t)$ contribute to the evolution of the source function $Q(K,t).$ In turn, the electron distribution $N(K; t)$ evolves following Equation~(\ref{eqn:fokker-planck}) in a way that depends upon the evolution of $Q(K,t)$. 
We show this evolution in Figure~\ref{fig:rxj1713evol}, again using the parameters in Table~\ref{tab:rxj1713fitparams} as a guide for physically realistic quantities.

Initially, the electron distribution approximately follows a power law at energies $K<K_c$, and a broad peaked distribution at higher energies. The shape of this distribution remains relatively unchanged while the overall number of accelerated electrons increases until $t=t_{\rm Sed}$. After the Sedov time, the SNR begins to slow down, reducing the maximum energy to which electrons are accelerated. As a result, the distribution of accelerated electrons begins to favor low energy electrons. At late times, the acceleration of low energy electrons decreases, and substantial energy loss through synchrotron and inverse Compton radiation leads to a ``break'' in the electron distribution. The overall electron distribution begins to decrease due to the transition from the Sedov phase to radiation phase at $t\approx 60t_{\rm Sed}$.

From the electron distributions, we can model the changes to the photon emission spectra over time. Throughout this evolution, high energy photon emission is dominated by inverse Compton scattering off of cosmic microwave radiation and the interstellar radiation field, including 
direct stellar radiation and stellar radiation absorbed and re-emitted by dust.  Low energy emission is dominated by synchrotron radiation. Bremsstrahlung plays only a subdominant role at high energies until very late times. Initially, as the total number of accelerated electrons increases, the emission likewise increases at all frequencies. After the Sedov time, the shift in electron distribution toward lower energies leads the emission spectrum to also shift to lower frequencies. Generally, reduction in the acceleration of accelerated electrons also leads the emission spectrum to decrease in overall intensity after the Sedov time. However, the increase of low energy electrons gives rise to the formation of a small bump in the X-ray spectrum around $t \approx 32 t_{\rm Sed}$. 

\begin{figure*}
    \includegraphics[width=\columnwidth]{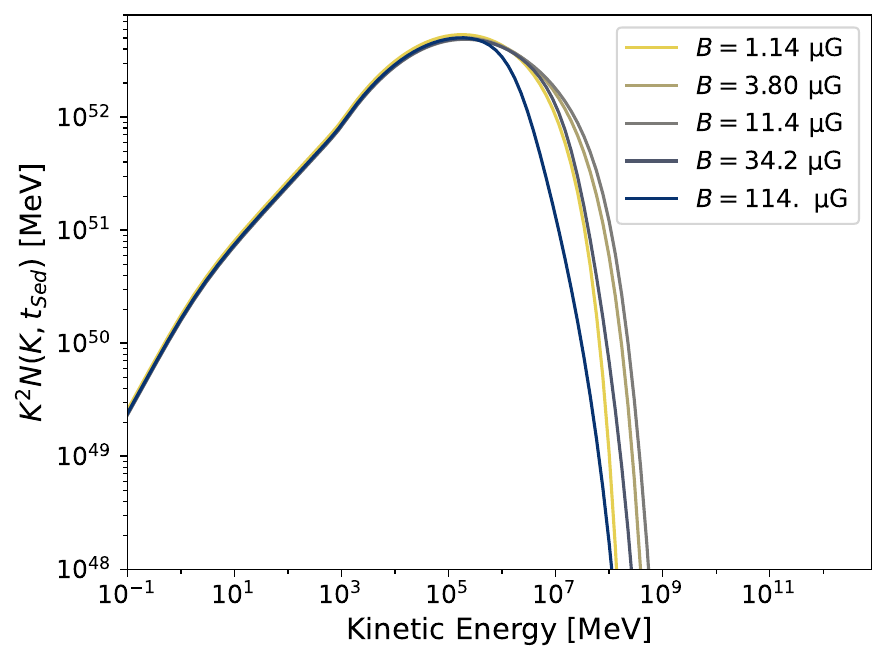}
    \includegraphics[width=\columnwidth]{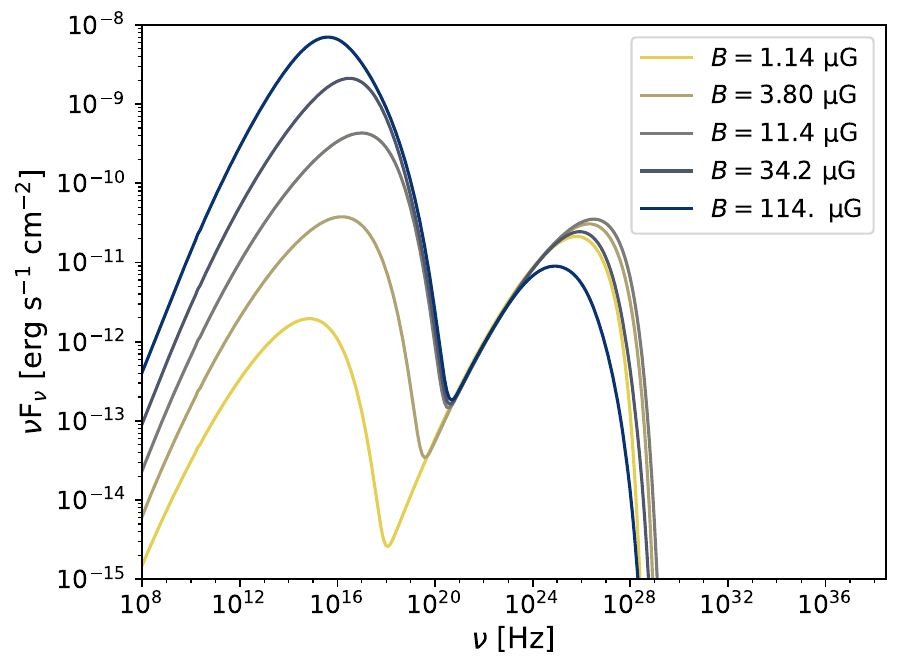}
    \caption{ Dependence of electron distribution (left) and photon spectrum (right) on the magnetic field $B$.}
    \label{fig:rxj1713_bcompare}
\end{figure*}

\subsection{Variation in Magnetic Field}

While observations can offer some constraints, the inherent complexity of the magnetic field in and around SNRs makes precise measurements difficult. With this in mind, we consider how altering the magnetic field affects the electron distribution and photon emission spectrum. In Figure~\ref{fig:rxj1713_bcompare}, we plot the electron distribution and SED resulting from modifying the model described in Table~\ref{tab:rxj1713fitparams} to have magnetic field strengths up to an order of magnitude greater or less than in Table~\ref{tab:rxj1713fitparams}. 

The electron distribution is affected by the magnetic field primarily through the electron maximum energy $K_{max}$, as described in Equations~(\ref{eqn:kmax1}) and (\ref{eqn:kmax2}). A weaker magnetic field results in electrons re-crossing the shock-front less frequently, and consequently at a given time early in the evolution of the SNR a weaker magnetic field leads $K_{max,1}$ to be lower resulting in a lower maximum electron energy overall. On the other hand, stronger magnetic fields increase the rate of synchrotron emission, leading $K_{max,2}$ to be lower and a lower maximum electron energy overall again. Both of these effects can be seen in Figure~\ref{fig:rxj1713_bcompare}, where the magnetic field in Table~\ref{tab:rxj1713fitparams} allows the least suppression of high-energy electrons of the magnetic fields shown here. When the magnetic field is weaker, the maximum energy is determined by $K_{max,1}$, which is lower than with the parameters in Table~\ref{tab:rxj1713fitparams}. When the magnetic field is stronger, the maximum energy is determined by $K_{max,2}$, which is also lower than with the parameters in Table~\ref{tab:rxj1713fitparams}.
Furthermore, the emission of synchrotron radiation reduces the number of high-energy electrons, and for stronger magnetic fields, this direct reduction of high-energy electrons is increased.

These effects on the electron distribution directly affect the photon spectra, as seen in the right side of Figure \ref{fig:rxj1713_bcompare}.  Additionally, it directly affects the synchrotron emission; the stronger magnetic fields create more synchrotron radiation than weaker magnetic fields.

\begin{figure*}
    \includegraphics[width=\columnwidth]{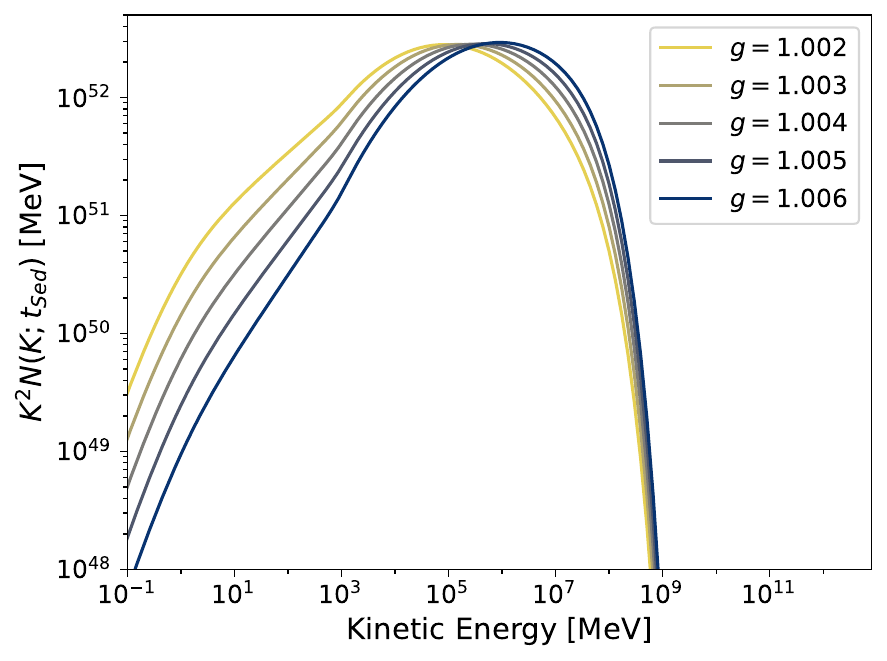}
    \includegraphics[width=\columnwidth]{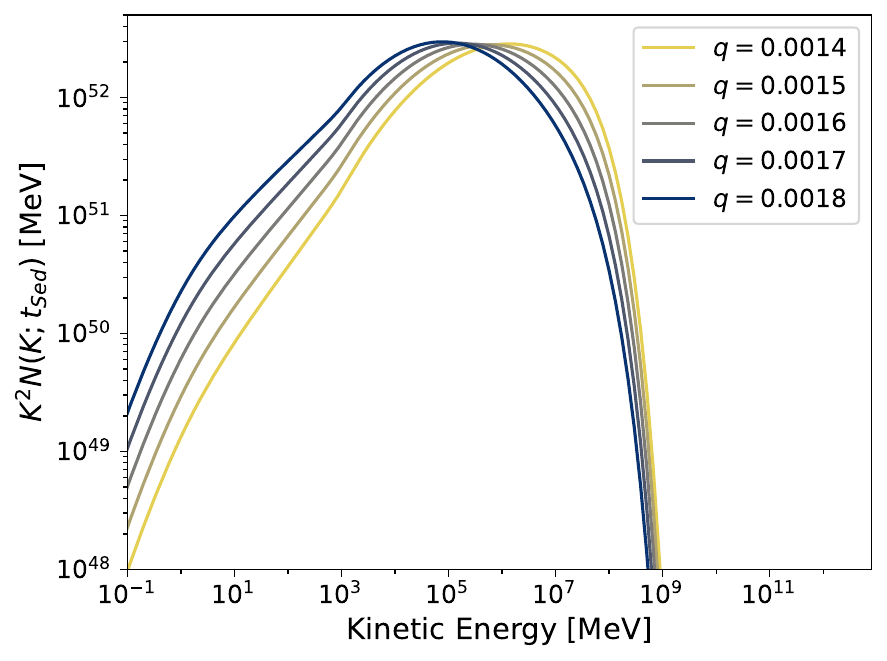}\\
    \includegraphics[width=\columnwidth]{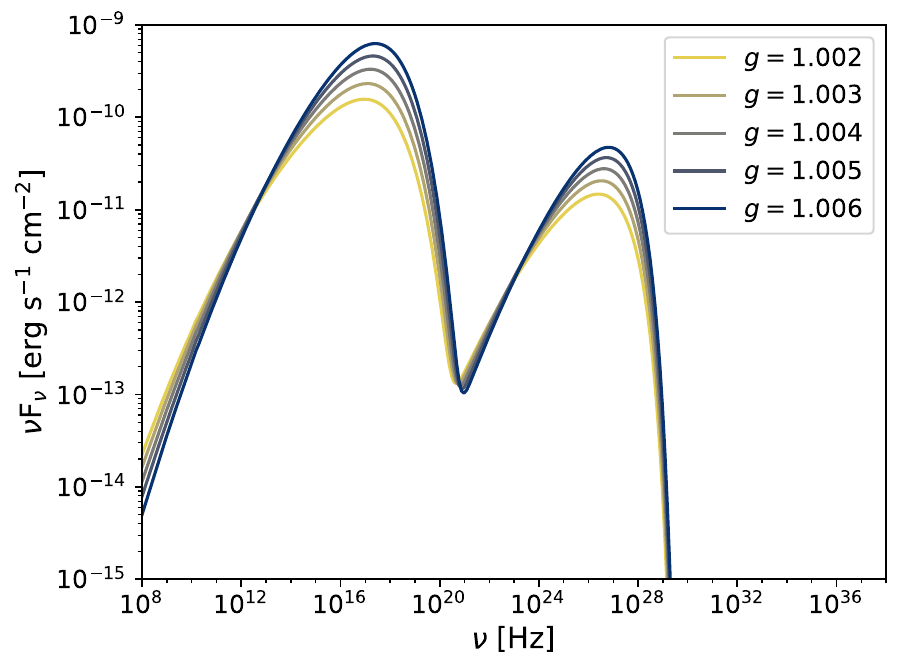}
    \includegraphics[width=\columnwidth]{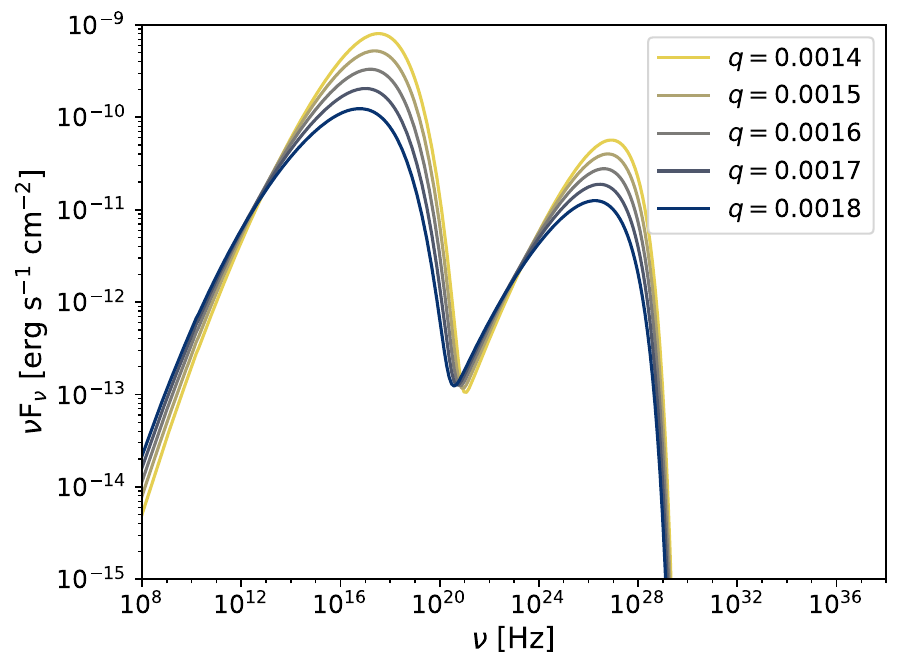}
    \caption{Dependence of electron distribution (top) and photon spectrum (bottom) on small changes in the $g$ (left) and $q$ (right) parameters around the values in Table~\protect{\ref{tab:rxj1713fitparams}} at the Sedov time ($t=t_{\rm Sed}$). 
    Even small changes in either $g$ or $q$ can produce significant differences in both the electron distribution and the photon spectrum.  
    }
    \label{fig:rxj1713_qgcompare}
\end{figure*}

\subsection{Variation in Parameters $g$ and $q$}

Here we explore the impact of changing the parameters $g$ and $q$ that define the log-parabola distribution on the electron distribution and photon emission spectra. We plot the results of these comparisons in Figure~\ref{fig:rxj1713_qgcompare}.  The time $t=t_{\rm Sed}$ is used as our fiducial time for this exploration.  The most notable observation is that even small changes in the values of $g$ and $q$ lead to significant changes in the electron distribution and to the emission spectra. Increasing the value of $q$ shifts the peak of the electron distribution to lower momenta while simultaneously flattening the power law at low energies. Because the exponential falloff of the electron distribution at high energies is determined independently of the acceleration model, changing $q$ does not change the asymptotic upper energy of the electron distribution, but because the peak shifts to lower energy, the shape of this falloff becomes gentler.
Increasing $g$, meanwhile, shifts the peak of the electron distribution to higher momenta while steepening the power-law at low energies. The effects of decreasing $q$ and increasing $g$ are similar but not fully degenerate. For example, increasing $g$ increases slightly the number of electrons at the peak of the distribution, while decreasing $q$ decreases slightly this number.

The effects of changing $g$ and $q$ on the emission spectra are less extreme, but still significant, particularly near the peaks of the spectra. Again, increasing $g$ and decreasing $q$ have comparable but not completely degenerate effects. In either case (increasing $g$ or decreasing $q$) leads emission near the peaks of the spectra to increase, while the emission at the trough between these peaks remains fairly constant. Furthermore, both increasing $g$ and decreasing $q$ cause a slight steepening of the low frequency spectrum. However, there is a slightly greater increase at the peak for the same amount of steepening at low frequencies when decreasing $q$ compared to increasing $g$. Ultimately, this means that simultaneously increasing $g$ and $q$ can allow the spectral peaks to remain at approximately the same intensity, but allow a fine control over the slope of the low-frequency spectrum. It is this control that allows us to improve upon previous one-zone fits to SNR spectra, like that of \rxj.

\section{Application to RX J1713.7$-$3946}
\label{RXJ1713}

The SNR \rxj\ (also known as G 347.3$-$0.5) is a well-studied young shell-type remnant, whose nonthermal emission was first observed in X-ray observations by ASCA in 1997 \citep{koyama97}. Subsequent observations have been made across a wide range of wavelengths including radio \citep[e.g.,][]{lazendic04}, infrared \citep[e.g.,][]{benjamin03, acero09}, X-ray \citep[e.g.,][]{koyama97,uchiyama03, cassam-chenai04}, GeV $\gamma$-ray \citep[e.g.,][]{abdo11}, and TeV $\gamma$-ray \citep[e.g.,][]{muraishi00, enomoto02,aharonian06}. \rxj\ has been identified as the remnant from a supernova observed in 393 C.E. by Chinese astronomers \citep{wang97}, suggesting an age of approximately 1630 years. Gas clouds in \rxj\ have been measured to be located at approximately $d=1~\kpc$ \citep{fukui03}. Furthermore, within the SNR a neutron star has been observed \citep{lazendic03} suggesting that \rxj\ was produced by a core collapse supernova, with studies of the chemical composition of the ejecta and the expansion rate of the SNR further suggesting that the supernova was type 1b/c of a comparatively low-mass star \citep{katsuda15}. The mass of the stellar progenitor has been estimated at $12-20~M_\odot$ \citep{cassam-chenai04, katsuda15}, so the mass ejected during the supernova can be estimated at greater than $5~M_\odot$ \citep{kennicutt84} with an explosion energy of $\sim 10^{51}~\erg$.  The lack of thermal X-ray lines indicates a CSM density $\la0.2\ \cm^{-3}$ \citep{slane99,ellison10}.  Given its low density, the Sedov timescale is expected to be large.  For instance, \citet{tsuji16} used models with Sedov timescale $t_{\rm Sed}\approx10^3$ --- $10^5$\ year, the latter times indicating that \rxj\ has not yet entered the Sedov phase. \citet{brose20} use $t_{\rm Sed}\approx10^3$\ year.
\rxj\ has an angular radius of 40\arcmin\ \citep{uchiyama03}, so that given its distance of $d=1\ \kpc$, it has a radius of $r_t\approx 10\ \pc$.

\begin{figure*}
    \includegraphics[width=\textwidth]{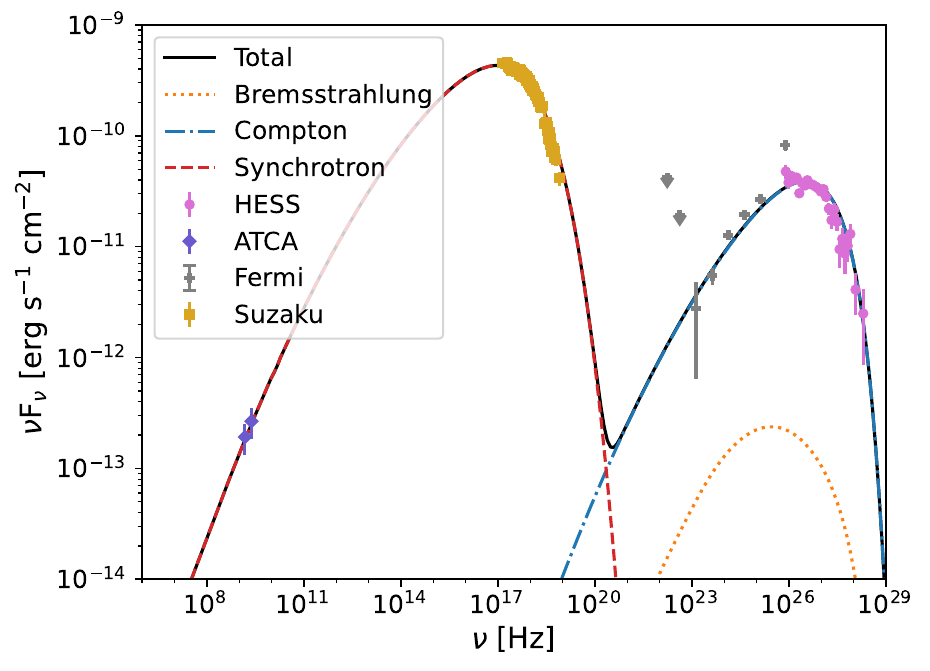}
    \caption{Model SED compared to observations of \rxj. HESS data are from \citet{aharonian06}, ATCA data from \citet{aharonian06,aharonian07}, Fermi data from \citet{abdollahi22}, and Suzaku data from \citet{tanaka08}. Contributions from bremsstrahlung (orange dotted curve), inverse Compton scattering of background radiation (blue dash-dotted curve), and synchrotron radiation (red dashed curve) are shown in addition to the total (black solid curve). Model details are provided in the main text, and model parameters are in Table~\protect{\ref{tab:rxj1713fitparams}}.
    }
    \label{fig:rxj1713fit}
\end{figure*}

Using these parameters as a baseline, we can find values of the source function parameters that suitably approximate the observed spectra of \rxj\ from multiple frequency ranges. To perform this parameter estimation, we use gamma-ray data from HESS \citep{aharonian06} and Fermi-LAT \citep{abdollahi22}, X-ray data from Suzaku \citep{tanaka08}, and radio data from ATCA \citep{aharonian06,aharonian07}. Previous fits to observed spectral data, using a power-law electron acceleration model, generally fail to simultaneously fit the X-ray/$\gamma$-ray emission and radio emission, in a one-zone model. By using a log-parabola acceleration model, we are able to much more easily describe both the high-energy X-ray and $\gamma$-ray data and the low-energy radio data simultaneously.
We plot a model which agrees with these data reasonably well in Figure~\ref{fig:rxj1713fit} and list the parameters of that model in Table~\ref{tab:rxj1713fitparams}.  

For our Compton-scattering calculations, the model we ultimately use consists of three blackbody spectra centered at temperatures of $T_{\rm CMB}=2.725~\Kelvin$ (cosmic microwave background or CMB),  $T_{\rm IR}=25~\Kelvin$ (infrared or IR), $T_{\rm opt}=5000~\Kelvin$ (optical). The energy densities in these 3 blackbodies are, respectively, $U_{\rm CMB}=2.6\times10^{-7}~\MeV\,\cm^{-3}$, $U_{\rm IR}=3.0\times10^{-7}~\MeV\,\cm^{-3}$, and $U_{\rm opt}=1.84\times10^{-7}~\MeV\,\cm^{-3}$. This is roughly consistent with the interstellar radiation fields found by \citet{porter06}.

We highlight the small value of $q=0.0016$ and the near-unity value of $g=1.004$. These imply only a small deviation from power law acceleration is sufficient to match observed data. 
However, even with this small value of $q$, the fact that it is non-zero has a significant effect on the electron distribution and the SED (see Figure~\ref{fig:rxj1713_qgcompare}, Section~\ref{Evolution}).  We note that at low energies in this model, almost none of the particles escape; the probability that particles will remain in the shock and get accelerated is $P_0=0.993$.  The radius at the present time from the model is roughly consistent with the observed value. 

\begin{figure}
    \includegraphics[width=\columnwidth]{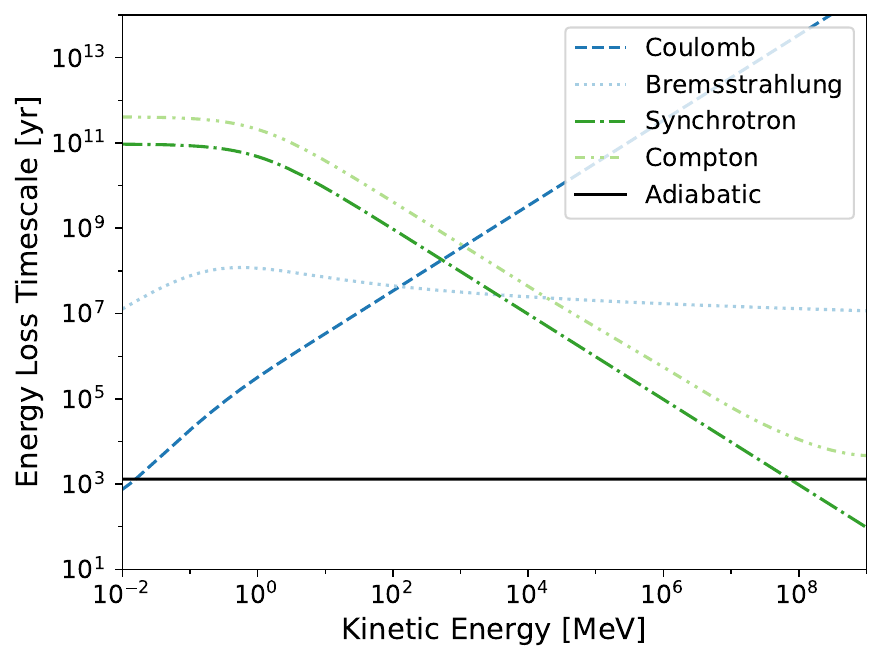}
    \caption{Comparison of electron energy loss timescales. For each mechanism by which electrons lose energy, we calculate an energy loss timescale as $t_{\rm cool}=|K/\dot{K}|$, evaluated for the model based on parameters in Table~\ref{tab:rxj1713fitparams} at time $t=t_{Sed}$. The adiabatic timescale (solid, black curve) has the shortest timescale at energies $K\lesssim10^8~\mathrm{MeV}$, while loss by synchrotron radiation dominates at higher energies. As the SNR evolves and its expansion slows, the timescale of the adiabatic cooling increases, so that at lower energies the Coulomb losses dominate.}
    \label{fig:energyloss}
\end{figure}

In Figure \ref{fig:energyloss} we have plotted the energy loss timescales, $t_{\rm cool}=|K/\dot{K}|$, where $\dot{K}$ is the energy loss rate.  The loss rates are computed following \citet{sturner97}.  We show the energy loss timescales for synchrotron, bremsstrahlung, Compton scattering, Coulomb interactions, and adiabatic losses from the expansion of the remnant.  All of these energy loss timescales are time-independent in our model except for the adiabatic losses.  At early times, in the free-expansion phase, adiabatic losses dominate.  These slowly decrease during the Sedov phase.  The radiative loss timescales are all longer than the age of the remnant, except for the Coulomb losses at low energies, and the synchrotron losses at high energies.  

\begin{figure}
    \includegraphics[width=\columnwidth]{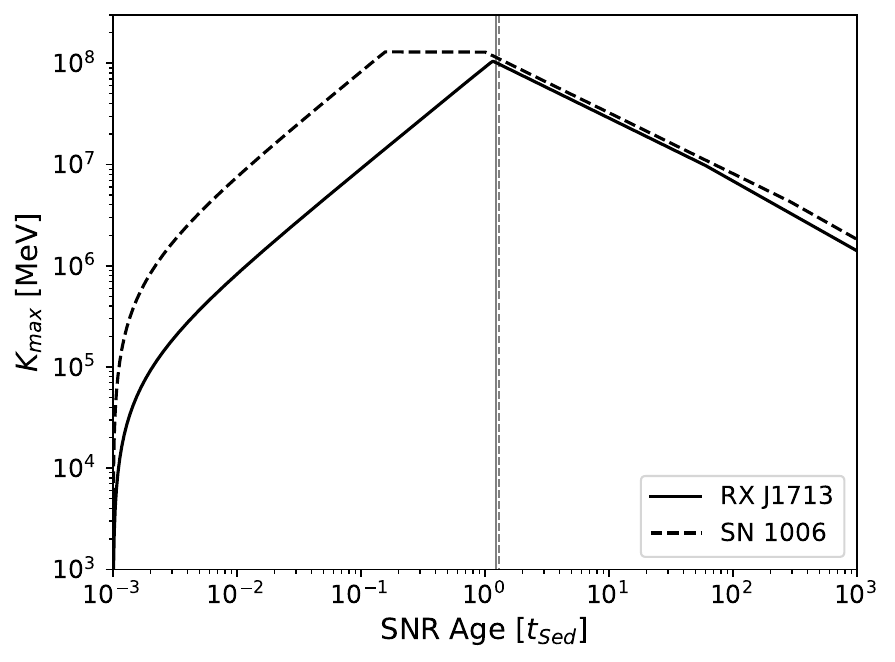}
    \caption{Maximum energy of electrons accelerated by an SNR as a function of the SNR's age. Two models that reflect the SNRs \rxj\ (solid line) and SN 1006 (dashed line) are shown. A thin gray vertical line shows the time at which each of these SNRs is observed. At early times, the maximum electron energy is limited by the age of the SNR, while at late times it is limited by energy loss to radiation exceeding the energy gain from shock acceleration. The transition point is the first kink in each curve.}
    \label{fig:kmax}
\end{figure}

In Figure~\ref{fig:kmax}, we plot the value of $K_{max}$ as a function of time for the model for \rxj. Initially, electrons cannot reach arbitrarily high energies because it would take longer than the age of the SNR for them to cross the shock front enough times to accelerate to those energies. As a result, $K_{max} = K_{max,1}$, and monotonically grows as the age of the SNR increases. At later times, the energy lost by accelerated electrons to radiation exceeds the energy gained through further acceleration, setting the maximum electron energy $K_{max} = K_{max,2}$. This maximum energy is constant until the Sedov time when the SNR begins to slow down and the energy gain through acceleration is reduced. 

\section{Application to SN 1006}\label{SN1006}

As with \rxj, we can use observations of SN 1006 (also known as G 327.6$+$14.6) to constrain the free parameters of the log-parabola model. SN 1006 was observed to occur in 1006 C.E. by multiple cultures \citep{katsuda17}. As such, it has an age of just over 1000 yrs. The remnant of SN 1006 was first observed by radio observations by \citep{gardner65}. No compact object has been identified within SN 1006, suggesting that its progenitor was a type Ia supernova, and consequently to have an ejected mass close to the Chandrasekhar mass $1.44~M_\odot$ and an initial energy of approximately $10^{51}~\erg$. Previous estimates for the mass of Fe ejected have been in tension with SN 1006 as a type Ia supernova, but recent analysis has resulted in a higher estimate, consistent with expectations for a type Ia supernova \citep{laming25}. The remnant has been localized to approximately $2.2~\kpc$ distance \citep[e.g.,][]{bandiera19}. It has an angular extent of $30'$ \citep{morlino10}, or a radius of $8.7~\pc$ at a distance of $2.2~\kpc$, and an angular expansion speed of $0.28''~\yr^{-1}$, corresponding to $\sim3\times10^8~\cm\,\s^{-1}$ \citep[e.g.,][]{katsuda09}.  A relatively low ISM density has been estimated; for instance $n_{ISM}\approx 0.03$ --- $0.1~\cm^{-3}$ by \citet{bandiera19}; or $n_{\rm ISM}\approx0.06~\cm^{-3}$ by \citet{berezhko12}.  This leads to a Sedov time for SN 1006 roughly equal to its current age. 

\begin{figure*}
    \includegraphics[width=\textwidth]{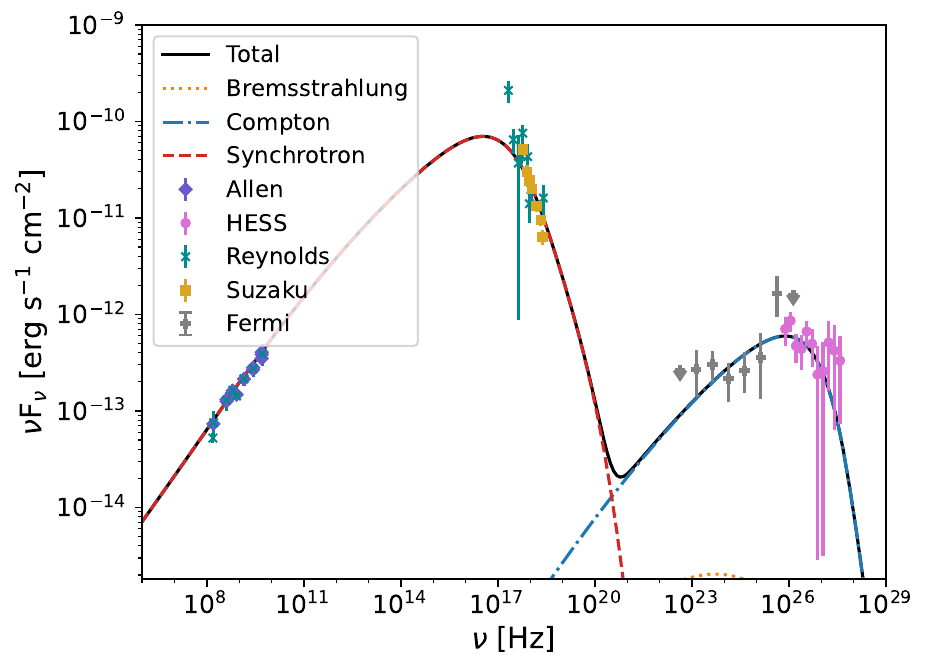}
    \caption{Similar to Figure~\protect{\ref{fig:rxj1713fit}},
     for the SN 1006 remnant. Radio data are taken from \citet{reynolds96, allen08}, X-ray data from \citet{reynolds96,acero10}, and $\gamma$ ray data from \citet{acero10, abdollahi22}, while curves have the same meanings as in Figure~\protect{\ref{fig:rxj1713fit}}. Model parameters are in Table~\protect{\ref{tab:snr1006fitparams}}.  
     }
    \label{fig:snr1006fit}
\end{figure*}

We use Fermi \citep{abdollahi22} and HESS \citep{acero10} $\g$ ray data, Suzaku X-ray data \citep{reynolds96,acero10}, and MOST \citep{allen08} and ASCA radio data \citep{reynolds96} to constrain our model for the SNR 1006 flux. Whereas \rxj\ proves challenging to fit with a power-law acceleration model, fitting SN 1006 with a power-law acceleration model has been significantly more successful. Therefore, using a log-parabola acceleration model to describe SN 1006 data allows us to explore how generalizable this model may be. Figure~\ref{fig:snr1006fit} shows a plot of one such model that approximates observed data reasonably well, and Table~\ref{tab:snr1006fitparams} lists the parameters used to produce this model. We model the background as a combination of four blackbody spectra centered at temperatures of $T_{\rm CMB}=2.725~\Kelvin$,  $T_{\rm IR}=25~\Kelvin$, $T_{\rm opt}=5000~\Kelvin$, The energy densities in these blackbodies are, respectively, $U_{\rm CMB}=2.6\times10^{-7}~\MeV\,\cm^{-3}$, $U_{\rm IR}=2.2\times10^{-7}~\MeV\,\cm^{-3}$, and $U_{\rm opt}=2.2\times10^{-7}~\MeV\,\cm^{-3}$.

The value $q$ is quite close to 0, indicating very little dependence of particle escape on energy.  However, as with \rxj, and as described above, this small change can have a significant effect on the SED.  As with \rxj, the probability the particles will be accelerated at low energies is nearly unity ($P_0=0.970$).  The radius and speed from the model are roughly consistent with the present values. 

As in the \rxj\ case, we note the particularly small value of $q=0.0016$ and the near-unity value of $g=0.9805$.
Although the value of $g<1$, because momentum can take any value $p>0$, there is still a momentum below which the probability would have an unphysical value. As such, we again impose a flat probability below the cutoff energy, taken to be $10^3$ MeV.

\begin{table}
\caption{Model parameters for SN 1006 remnant.} 
\centering
\begin{tabular}{l c c}
\hline
\hline
Parameter & Symbol & Model\\
\hline
\multicolumn{3}{c}{Parameters fixed by observation}\\
\hline
Age [yr] & $t$ & 1000\\
Distance [kpc] & $d$ & $2.2$\\
\hline
\multicolumn{3}{c}{Free parameters constrained by observation}\\
\hline
Initial Mass [M$_\odot$] & M$_{ej}$ & 1.4\\
Initial Energy [$10^{51}$ erg] & $E_0$ & 1.2\\
ICM Density [cm$^{-3}$] & $n_{ISM}$ & 0.025 \\
Magnetic field [$\mu$G] & $B$ & 40\\
\hline
\multicolumn{3}{c}{Free parameters}\\
\hline
Acceleration scaling parameter & $q$ & 0.0016\\
Acceleration normalization [MeV/c] & $g$ & 0.9805\\
Log-Parabola electron energy [MeV] & $K_0$ & $10$\\
Log-Parabola cutoff energy [Mev] & $K_c$ & $10^{3}$\\
Electron acceleration efficiency & $\eta_e$ & $2.0\times10^{-4}$\\
\hline
\multicolumn{3}{c}{Derived parameters}\\
\hline
Low-energy probability of acceleration & $P_0$ & 0.970 \\
Initial velocity [cm s$^{-1}$] & $v_0$ & $9.3\times 10^{8}$\\
Sedov time [yr] & $t_{\rm Sed}$ & 767\\
Present radius [pc] & $R_t$ & 12.2\\
Present velocity [cm s$^{-1}$] & $v_t$ & $7.9\times 10^8$\\
\hline
\end{tabular}
\label{tab:snr1006fitparams}
\end{table}

The maximum energy of electrons accelerated by SN 1006 can be seen in Figure~\ref{fig:kmax}. Like the maximum electron energy from \rxj, at early times, this maximum is determined by the age of the SNR, and at late times by the rate of synchrotron emission. However, the time at which this transition occurs happens somewhat before the Sedov time for SN 1006, and approximately at the Sedov time for \rxj. The higher initial velocity of SN 1006 means that electrons encounter the shock front more frequently, allowing the maximum electron energy to increase more rapidly over time.

\section{Discussion}
\label{Discussion}

An electron acceleration model characterized by a log-parabola model can naturally agree with observations of SNR emission spectra from radio to TeV $\gamma$ waves. Moreover, we are able to produce such log-parabola models that agree with data of both \rxj, which a single-zone power-law acceleration model struggles to fit naturally, and SN 1006 which has been fit well with a single-zone power-law acceleration model. While having three extra free parameters in the log-parabola model (compared to the power-law model) makes it perhaps unsurprising that this model is more adaptable to different SNRs, an intriguing coincidence also appears when modeling these two SNRs. Namely, fixing the values of $K_0=10~\rm{MeV}$ and $K_c=10^3~\rm{MeV}$, the values of $q$ and $g$ are comparable in both cases. In particular, the value of $q\approx 0.0016$ is favored in both SNRs. The value of $g$ shows significantly more variation, but is still around unity in both cases. It would be interesting to explore whether this pattern holds in additional SNRs.

The low value of $q$ in particular suggests that the probability of accelerating (or conversely the probability of escaping the shock-front) is only slightly energy-dependent, and then only above a cutoff energy. Although the assumption of energy-independent escape made by \citet{bell78_paper1} may seem to be justified based on the low value of $q$, even the small value of $q$ can in fact make a significant difference in the electron distribution (see Figure \ref{fig:rxj1713_qgcompare}).  Some energy-dependence at high-energy does seem to be required.

\begin{figure}
\includegraphics[width=\columnwidth]{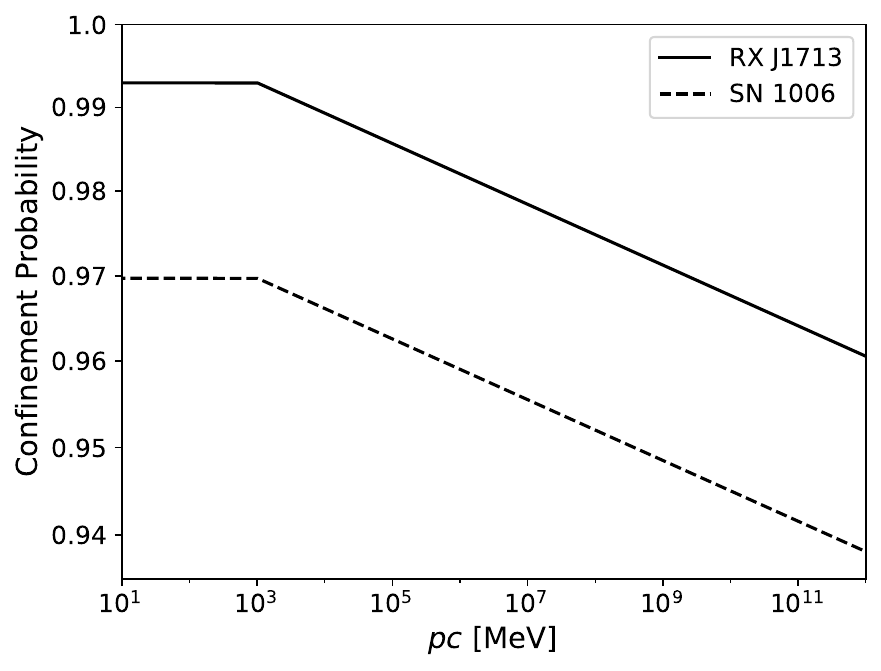}
\caption{Probability that an electron will \emph{not} escape the SNR shock region, and will be further accelerated to higher energies. The solid curve represents the model used to describe \rxj, while the dashed curve represents the model used to describe SN 1006. The low value of $q=0.0016$ in both models implies that electrons will have a high probability of remaining bound within the shock region even if they have a high energy.}
\label{fig:prob}
\end{figure}

We plot the probability as a function of momentum for both SNR models we consider in Figure~\ref{fig:prob}; for the \rxj\ model the probability of remaining confined is greater than 99\% for electrons with momenta $pc\lesssim 10^4~\MeV$ and greater than 97\% for electrons with $pc\lesssim 10^{10}~\MeV$ and for the SN 1006 model the probability for remaining confined is about 97\% for $pc\lesssim 10^3~\MeV$ and greater than 95\% for $pc<10^9~\MeV$. Physically this implies that almost all electrons are scattered back into the shock region by interactions with the shock front, with only a very small fraction being scattered out. Furthermore, confinement by magnetic fields near or within the shock front are strong enough that only with significantly high energies can electrons consistently escape.  

\begin{acknowledgements}
We are grateful to the referee for comments that have improved this paper. The authors are supported by the Office of Naval Research.
\end{acknowledgements}

\appendix

\section{Probability of Acceleration}

In our analysis, we use a broken power law to describe the probability of acceleration by a SNR shock front. However, a more physically realistic model of the probability would smoothly transition between these asymptotic behaviors. \citet{massaro04} suggest a functional form for the probability that achieves this: $P=g(1+(p/p_c)^q)^{-1}$.  
However, because the value of $q$ that we use is so close to 0, this function does not have the desired behavior over reasonable values of $p$. To correct for this, we consider here a slightly different function with the same asymptotic behavior at low and high momenta: 
\begin{equation}
    P=\frac{g}{(1+p/p_c)^q}.
\end{equation}
This function for the probability does not lend itself to an explicit function for $Q(>pc)$, or consequently of $Q(pc)$, where the broken power law does.

\begin{figure*}
    \includegraphics[width=0.498\textwidth]{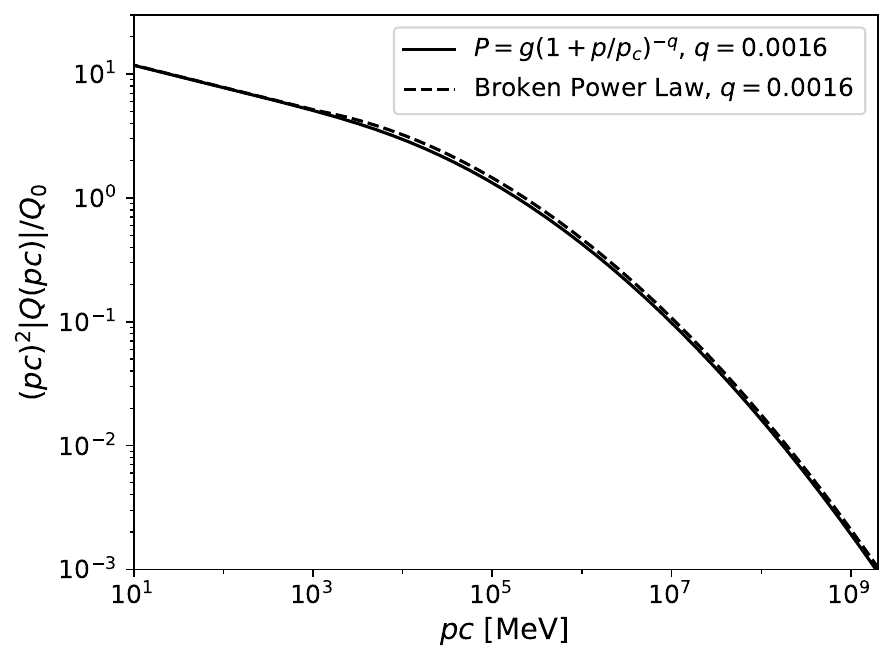}
    \includegraphics[width=0.498\textwidth]{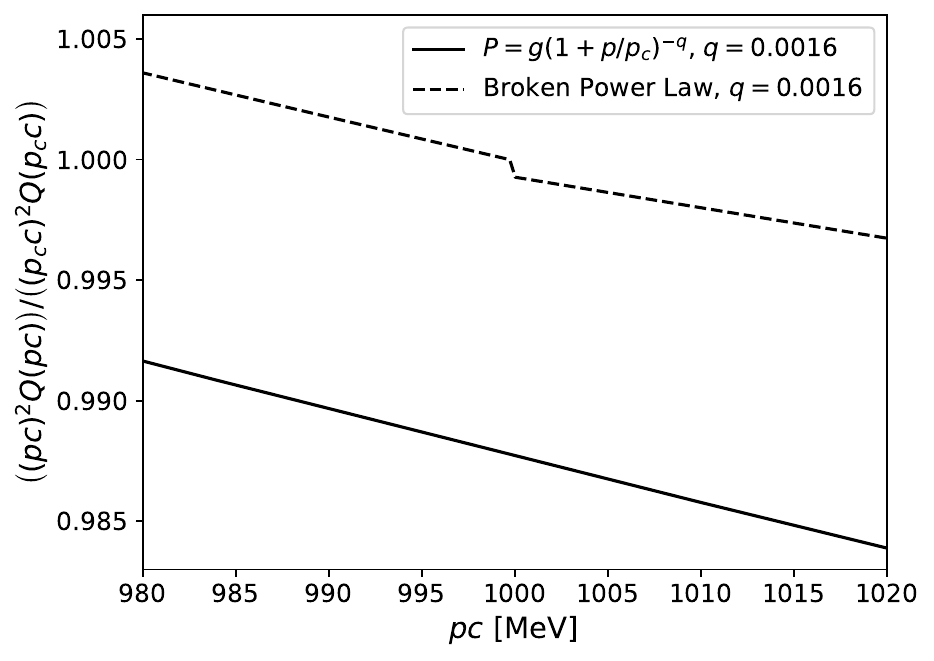}\\
    \includegraphics[width=0.498\textwidth]{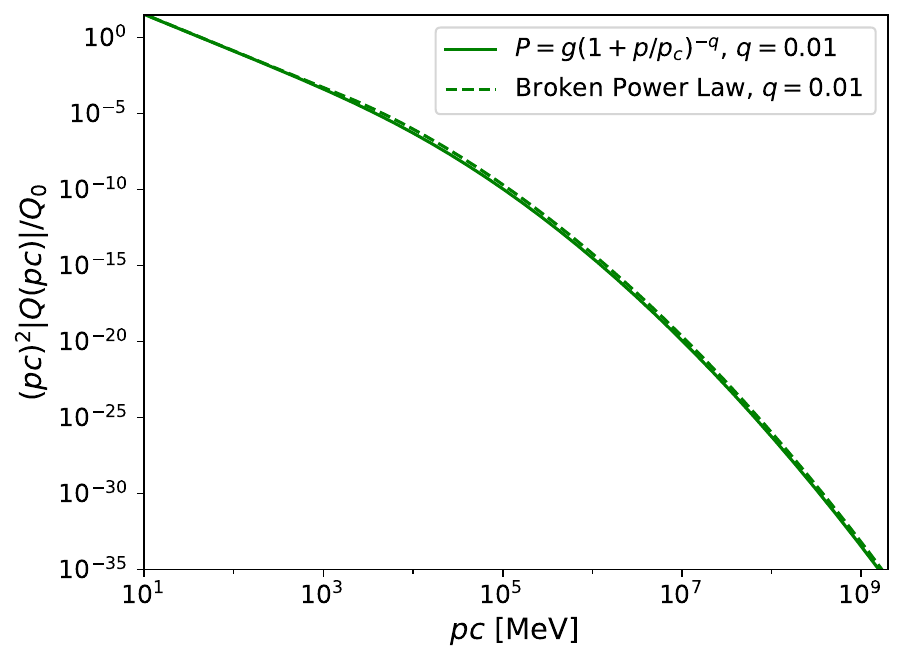}
    \includegraphics[width=0.498\textwidth]{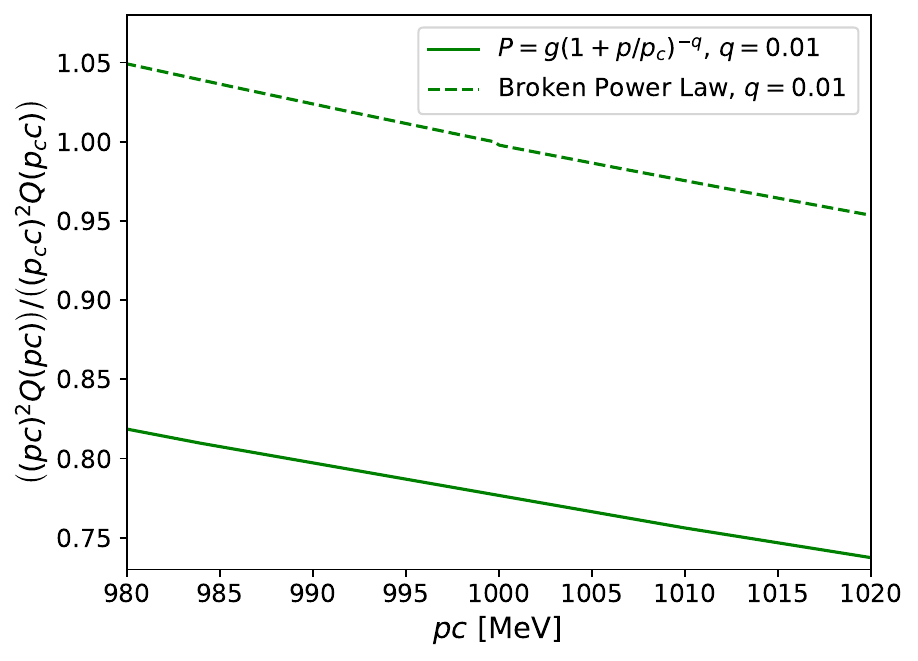}
    \caption{Comparison of the electron source functions $Q(pc)$ that arise from the broken power law probability distribution we use throughout this work and the probability distribution $P= g(1+p/p_c)^{-q}$, suggested by \protect{\citet{massaro04}}. 
    The left panels show the source function multiplied by $(pc)^2$, normalized so that the number of electrons with momentum $p_0$ is $Q_0$. Dashed curves represent the source functions derived from the broken power law probability distribution, while solid curves represent that from the continuous probability distribution. In each case, we use parameters from the SN1006 remnant, as described in Table~\ref{tab:snr1006fitparams}, including $p_cc = 10^3~\mathrm{MeV}$. In addition, we also plot (bottom) a comparison assuming a larger value of $q=0.01$. The right plots show the region around the $pc=p_cc=1000~\MeV$, to highlight the discontinuous nature of the broken-power-law acceleration probability. In these plots, each curve shows the ratio of $(pc)^2Q(pc)/Q_0$ to its value as the momentum approaches $p_c$ from the left ($p\rightarrow {p_c}^-$). That is, the value of this ratio will be 1 for both broken-power-law curves on the left side of the discontinuity. 
    }
    \label{fig:probmodelcompare}
\end{figure*}

We can numerically compute $Q(>pc)$ as
\begin{equation}
    Q(>pc) = \prod_{i=0}^{n-1} \frac{g}{\left(1+\frac{p_0}{p_c}\varepsilon^i\right)^q},
\end{equation}
where $p_i = p_0\varepsilon^i$. We use a finite differencing method to approximate $Q(pc)=dQ(>pc)/d(pc)\approx \Delta Q(>pc)/\Delta(pc)$.  We then compare the result to the algebraic results derived from the broken power law probability, and show the results in Figure~\ref{fig:probmodelcompare}. For this comparison, we consider the broken-power-law acceleration probability to be based on the model described in Table~\ref{tab:snr1006fitparams}. For the sake of comparison, we also consider a case with the larger value of $q = 0.01$ and all other parameters held the same. Comparing to the results from the right side of Figure~\ref{fig:rxj1713_qgcompare} suggests that this change from $q=0.0016$ to $q=0.01$ will have a substantial effect on the particle acceleration.

When comparing the source functions from the broken-power-law probability distribution and the continuous probability distribution, we note first that they are quite similar, especially for smaller $q$. For $q=0.0016$, the difference between these two source functions is $\mathcal{O}(1\%)$, while for the $q=0.01$ case, this increases to $\mathcal{O}(20\%)$. 

We note that the difference between the two acceleration models is only noticeable around and after $pc=p_cc=10^3~\MeV$. This is especially true in the $q=0.0016$ case, which is consistent with the probability distributions deviating in only a small region around this momentum. When $q=0.01$, we can see the difference between the two acceleration models beginning to appear at a lower momentum, which is consistent with the probability distribution showing marked differences over a wider range of momenta around $p=p_c$.

Finally, we note the discontinuity in the source function derived from the broken-power-law probability distribution, as shown in the right plot of Figure~\ref{fig:probmodelcompare}. From Section~\ref{Formalism}, we can derive that the ratio of the values of the source function when approached from the left and right sides is
\begin{equation}
    \frac{\lim_{p\rightarrow {p_c}^+}Q(pc)}{\lim_{p\rightarrow {p_c}^-} Q(pc)} = \frac{\left|\frac{\ln{g}}{\ln\varepsilon}-q\frac{\ln{p_cc}}{\ln\varepsilon} +\frac{q}{2}\right|}{\left|\frac{\ln{g}}{\ln\varepsilon}-q\frac{\ln{p_cc}}{\ln\varepsilon}\right|}.
\end{equation}
For the small values of $q$ we use, then the discontinuity is very small. However, for larger values of $q$, there is a higher probability an electron will escape being confined by the SNR shock front, so the overall magnitude of the source function is smaller. As a consequence, the difference between the left and right limits across the discontinuity will be larger for higher $q$, but the ratio between the right and left limits will be smaller. It is this latter quantity which is plotted in the right panels of Figure~\ref{fig:probmodelcompare}. While the discontinuity in the ratio in the $q=0.01$ case is larger than in the $q=0.0016$ case, the difference between the \citet{massaro04}-based model and the broken power law is also significantly larger.

\bibliography{SNR_reference}{}
\bibliographystyle{aasjournalv7}

\end{document}